\documentclass{article}

\usepackage{arxiv}

\usepackage{hyperref}       
\usepackage{url}            
\usepackage{booktabs}       
\usepackage{amsfonts}       
\usepackage{nicefrac}       
\usepackage{microtype}      
\usepackage{lipsum}
\usepackage{graphicx}
\usepackage{multirow}%
\usepackage{amsmath,amssymb,amsfonts}%
\usepackage{amsthm}%
\usepackage{mathrsfs}%
\usepackage[title]{appendix}%
\usepackage{xcolor}%
\usepackage{textcomp}%
\usepackage{manyfoot}%
\usepackage{booktabs}%
\usepackage{algorithm}%
\usepackage{algorithmicx}%
\usepackage{algpseudocode}%
\usepackage{listings}%
\usepackage{placeins}
\graphicspath{ {./images/} }

\title{Evaluating Large Language Models in Code Generation: INFINITE Methodology for Defining the Inference Index}

\author{
 Nicholas Christakis \\
  Institute for Advanced Modeling and Simulation\\
  University of Nicosia\\
  Nicosia, CY-2417, Cyprus \\
  \texttt{nchristakis@tem.uoc.gr} \\
   \And
 Dimitris Drikakis \\
  Institute for Advanced Modeling and Simulation\\
  University of Nicosia\\
  Nicosia, CY-2417, Cyprus \\
  \texttt{drikakis.d@unic.ac.cy} \\
}

\begin{document}
\maketitle
\noindent 
\begin{abstract}
This study introduces a new methodology for an Inference Index (InI), called INFerence INdex In Testing model Effectiveness methodology (INFINITE), aiming to evaluate the performance of Large Language Models (LLMs) in code generation tasks. The InI index provides a comprehensive assessment focusing on three key components: efficiency, consistency, and accuracy. This approach encapsulates time-based efficiency, response quality, and the stability of model outputs, offering a thorough understanding of LLM performance beyond traditional accuracy metrics. We applied this methodology to compare OpenAI's GPT-4o (GPT), OpenAI-o1 pro (OAI1), and OpenAI-o3 mini-high (OAI3) in generating Python code for the Long-Short-Term-Memory (LSTM) model to forecast meteorological variables such as temperature, relative humidity and wind velocity. Our findings demonstrate that GPT outperforms OAI1 and performs comparably to OAI3 regarding accuracy and workflow efficiency. The study reveals that LLM-assisted code generation can produce results similar to expert-designed models with effective prompting and refinement. GPT's performance advantage highlights the benefits of widespread use and user feedback.
\end{abstract}


\section{Introduction}\label{intro}

The rapid advancement of Large Language Models (LLMs) \cite{openai2022optimizing, openai2023gpt4} brings significant benefits across various applications. One area increasingly attracting the scientific community's interest is using LLMs for algorithmic coding. Writing computer code requires considerable time and effort, and leveraging artificial intelligence (AI) tools to expedite this process can help developers and researchers accelerate scientific discoveries in diverse fields and technological applications. Therefore, developing indicators for assessment and ensuring careful AI model validation and experimentation is essential. This approach will guide progress in the field and prevent misleading hype.

The Generative Pre-trained Transformers (GPT) series \cite{openai2022optimizing, openai2023gpt4} is utilizing the Transformer architecture \cite{translation1}, which relies on the self-attention mechanism. Transformers have found applications in various domains, including language translation \cite{translation1, translation2}, speech recognition \cite{speech1, speech2}, language modelling \cite{bertpaper}, and in scientific and engineering fields such as fluid dynamics \cite{Drikakis:2024_ML3, Drikakis:2024_ML2}. These models consist of layers of multi-head attention and position-wise feed-forward networks. They undergo a pre-training phase with large-scale text data in an unsupervised manner, followed by fine-tuning for specific tasks. One of the standout features of LLMs is their exceptional performance across a range of Natural Language Processing (NLP) tasks, including machine translation, question-answering, and text generation. This versatility allows them to handle numerous functions without needing specialized modifications.

However, several challenges persist in the development of Transformers and LLMs. These include biases in training data that can lead to biased outputs, as well as complexity, which may hinder interpretability and make modification difficult \cite{huang2023bias, NEURIPS2023_ae9500c4, 10.1145/3637528.3671458}. Furthermore, creating specialised code models, such as forecasting models for time series data, requires detailed domain knowledge to design optimal architectures and effectively tune parameters \cite{jin2023time, yu2023temporal}. Additionally, scalability can be an issue, as expanding to larger datasets or more complex tasks can present challenges \cite{zhang2024scaling}. All these factors may result in unintended outputs, while the black-box nature of LLMs complicates efforts to comprehend the reasoning behind their decisions. \cite{ajwani2024llm}. 

A particularly challenging area for LLMs is code generation. Unlike natural language, code necessitates strict adherence to syntax, logic, and context, making it difficult for LLMs to consistently produce correct and functional code \cite{nam2024using}. The absence of context awareness in multi-step programming tasks means that LLMs may produce code that functions in isolation but fails when incorporated into larger programs, often resulting in bugs or errors \cite{liu2024exploring}. Moreover, LLMs face challenges with debugging and validation because they cannot reason through errors or test their code in real time, leading to undetected mistakes \cite{tian2023chatgpt, liu2024refining}. While techniques like Chain of Thought (CoT) reasoning offer some improvement in transparency \cite{NEURIPS2022_8bb0d291, zhang2022automatic, jin2024impact}, LLMs still face significant hurdles in generating reliable, optimized code, often requiring human oversight to ensure correctness.

In AI literature, the process of running a trained model on new input data to produce an output is referred to as \textbf{inference} \cite{8675201, 10.1145/3183713.3183751, creswell2022selection, pmlr-v202-sheng23a}. To evaluate the performance and efficiency of inference, several key factors relating to the various stages of the process have to be taken into consideration and quantified. These factors will be discussed in the following section. Various inference models have been proposed to assess the performance of AI tools \cite{chitty2024llm, lukasik2024metric, he2024uellm}, but there is no standard metric for evaluating the effectiveness of inferences across tasks and applications
\cite{MARSILI20221, PSAROS2023111902}.  This study introduces a new Inference Index (InI) methodology for assessing code-writing with LLMs, labeled the INFerence INdex In Testing model Effectiveness (INFINITE). This methodology differs from other approaches due to its unique holistic approach to analyzing and quantifying the effects of key factors that impact the model's performance. This multidimensional assessment provides a richer picture of inference effectiveness and offers actionable insights into technical and operational aspects of deploying LLMs in code-generation tasks. Utilizing the INFINITE framework, we assess the coding abilities of different LLMs \cite{openai2022optimizing, openai2023gpt4}. We focus on a specific example of forecasting time series of meteorological variables, comparing ChatGPT-4o (GPT), OpenAI-o1 pro (OAI1), and OpenAI-o3 mini-high (OAI3) frameworks regarding their ability to generate and run a functioning long short-term memory (LSTM) model \cite{hochreiter1997long} when requested to do so. The results of the generated models are compared to the observations and predictions of an LSTM model developed by the authors (LSTM-H).

The paper is organized as follows: Section~\ref{sec:inf} introduces the INFINITE methodology. Section~\ref{sec:lstm} briefly introduces the LSTM model. In Section~\ref{sec:data}, an overview of the data set used in this study is provided. In Section~\ref{sec:preds}, the results from the trials and tests of AI models are presented, including InI calculations for each of the three frameworks. Finally, Section~\ref{sec:conclusions} reviews the findings and explores potential future directions.

\section{Inference and the INFINITE Methodology}
\label{sec:inf}
To develop the INFINITE methodology for assessing AI-generated codes, we will discuss key concepts that facilitate problem understanding and simplify the methodology and solution procedure. We define various questions (attempts) that will guide the LLM towards the desired answer. Whenever a question is posed, whether the same or a different one, it is treated as a single query. Following a query, the response may either be a "Server Busy" (SB) response or an answer, whether desirable or not. For example, if the same question is asked five times, with the first four yielding an SB response and the fifth providing any answer (desirable or otherwise), this counts as one attempt and five queries. If a non-satisfactory answer is received, a new question—considered a new attempt—will be posed. To calculate the average response time per query ($ARTpQ$) of a framework, we calculate the mean of the total response times, where response time ($RT$) is defined as the duration between a query and the framework's response:
\begin{equation}
ARTpQ = \frac{\Sigma_{i=1}^{N}RT}{N}
\label{eq:art}
\end{equation}
where $N$ is the total number of queries. 

When evaluating the performance of LLMs, including the OpenAI family frameworks, it is crucial to assess all intermediate steps in the process "from the initial question to the desired outcome." Therefore, we should consider and quantify factors such as the adequacy of computational resources needed for output generation, the time required to process each query and provide a response, the number of queries necessary to obtain the desired answer, and the quality of the responses. As previously mentioned, no standard metric for assessing inference effectiveness exists in the existing literature. Some attempts have shown promise, particularly causal inference, in improving the predictive accuracy, fairness, robustness, and explainability of LLMs \cite{liu2024largelanguagemodelscausal}. Causal inference evaluates assumptions, study designs, and estimation strategies to draw causal conclusions from data. It originates from three frameworks: potential outcomes, causal graphs, and structural equations. Potential outcomes concentrate on estimating causal effects, whereas graphical models—especially those based on Pearl’s causal graphs—demonstrate causal pathways using directed acyclic graphs (DAGs) to illustrate conditional independence among variables \cite{, pearl1998graphs,pearl2009causality, zeng2022surveycausalinferenceframeworks}. However, for code-generating tasks, causality is linked to the accuracy of results, particularly regarding whether a change in a model (for instance, switching from manually written code to AI-generated code) causes a genuine improvement in accuracy \cite{ji2023benchmarking}.

It is widely accepted that there are three major components impacting inference performance: \textbf{Efficiency}, \textbf{Consistency} and \textbf{Accuracy} \cite{vu2021scnet, mitchell2022enhancing, 10769165}. By \textbf{Efficiency}, we refer to how quickly the model processes and returns responses. It is influenced by (a) Response Time (i.e., the time taken for the model to generate a response after receiving a query); (b) Server Busy Rate (SBR, i.e., the proportion of queries that cannot be processed due to system overload or resource constraints); By \textbf{Consistency} we refer to how "clever" and well-trained the model is, to be able to produce satisfactory responses with as few attempts as possible. A key indicator is the number of attempts by the user until an acceptable response is obtained. By \textbf{Accuracy}, we refer to the model's reliability in providing a correct answer or making accurate predictions. Numerical predictions may be quantified through a mean relative error, which expresses how closely predicted values match the actual (ground truth) values.

To determine and quantify the corresponding parameters that allow for the assessment of these components influencing inference, we propose the \textbf{INFINITE} methodology (INFerence INdex In Testing model Effectiveness) along with the introduction of an Inference Index (InI). The steps of the methodology are as follows:
\begin{itemize}
    \item Firstly, we identify the key parameters involved in submitting a query, allowing us to measure Efficiency, Consistency, and Accuracy.
    \item As a second step, we develop indices ranging from 0 to 1, which define the performance of each individual factor separately; 0 represents the poorest performance, while 1 indicates excellent performance.
    \item. Finally, InI is calculated by weighting the average of individual indices. 
\end{itemize}

\begin{figure}[h]
    \centering
    \includegraphics[width=1.0\textwidth]{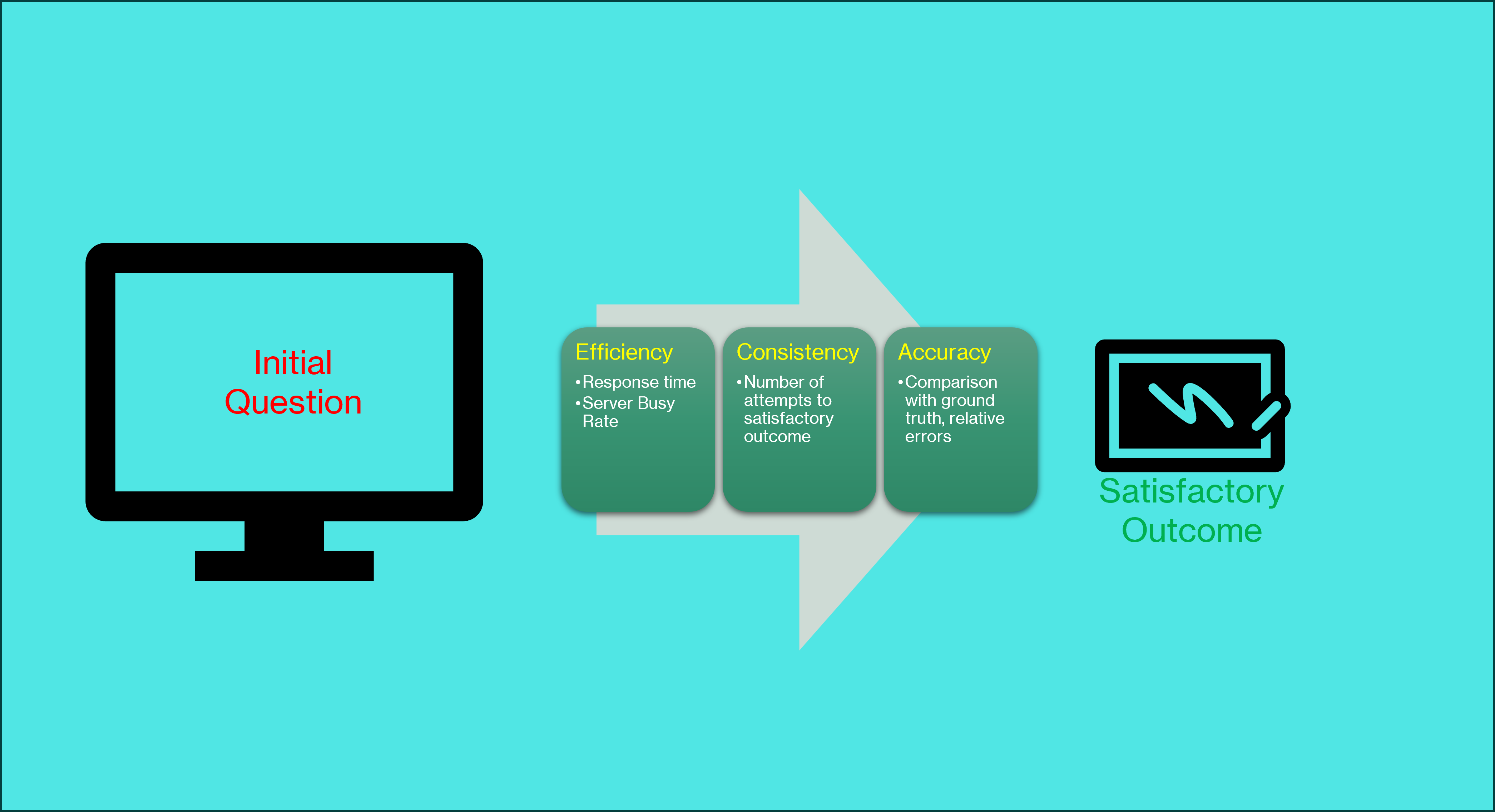}
    \caption{Major components impacting inference and their corresponding parameters that may be recorded while executing a code-generating task.}
    \label{fig:inf}
\end{figure}

For the first step of the methodology, key parameters that can be recorded during the execution of a code-generating task include the number of different questions (attempts) we must ask the model until the correct code is produced, the number of queries that need to be submitted each time (whether an SB response or an answer is returned), response times between a query and the response from the framework, SB responses, and relative errors, such as Mean Absolute Percentage Error (MAPE), between generated model predictions and ground truth. The SB responses, total number of queries, and response times are related to Efficiency. The number of questions posed until a satisfactory answer is achieved is tied to Consistency. The relative prediction error corresponds to Accuracy. In Figure~\ref{fig:inf}, we visually present the three major components and their corresponding parameters.

As we focus on code-generating tasks, we believe that metrics such as BLEU \cite{papineni2002bleu} or codeBLEU \cite{ren2020codebleu} are not well-suited for evaluating the accuracy of these tasks; both metrics measure the number of contiguous sequences of words (or tokens) from a generated text that match those in a reference text, rather than assessing the degree of deviation in code correctness. For this reason, we will only include MAPE in the accuracy assessment.  

As a second step, we offer suggestions for the indices that should be formulated to calculate InI. All index range from 1 (excellent performance) to 0 (poorest performance). First, we should consider both the SB rate and the $ARTpQ$ for Efficiency. Therefore, two indices will be formulated. We propose for the SB rate to be represented through the index $E_{SBR}$:

\begin{equation}
    E_{SBR} = 1 - \frac{\text{Number of SB responses}}{\text{Total queries}}
    \label{eq:esbr}
\end{equation}
The greater the number of SB responses the framework returns, the poorer its performance becomes. 

To account for the $ARTpQ$, we will define two threshold values of average response time, $ART_1$ and $ART_2$, where $ART_1 < ART_2$. Then, we may define index $E_{ART}$, such that the performance is excellent, average or poor:

\begin{equation}
    E_{ART} =
\begin{cases} 
1, & \text{if } \text{ARTpQ} \leq ART_1 \quad \text{(Excellent performance)} \\
0.5, & \text{if } ART_1 < \text{ARTpQ} < ART_2 \\
     & \quad \text{(Average performance)} \\
0, & \text{if } \text{ARTpQ} \geq ART_2 \quad \text{(Poor performance)}
\end{cases}
\label{eq:eart}
\end{equation}
The values for $ART_1$ and $ART_2$ depend on the user and the nature of the problem. We suggest reasonable values $ART_1=10 s$ and $ART_2 = 30 s$.

Then, an index $E$ for Efficiency may be derived from averaging the two indices. $E_{SBR}$ and $E_{ART}$:
\begin{equation}
    E = \frac{E_{SBR}+E_{ART}}{2}
    \label{eq:e}
\end{equation}

To construct a well-balanced decay function for the Consistency index $C$, we introduce $ln(Q)$ instead of simply using $Q$, where $Q$ represents the total number of attempts (i.e. questions asked). The main reason for this choice is that $ln(Q)$ grows significantly more slowly than $Q$, ensuring that the decay towards zero is gradual rather than abrupt. Given that $Q$ is always a positive integer starting from 1, the optimal scenario occurs when $Q=1$, signifying the problem is resolved with a single query. As $Q$ increases, the index should approach zero, reflecting the growing effort required to reach a conclusion. However, this decay should not be excessively rapid, as different problems exhibit varying complexity and ambiguity. To account for this variability, we introduce a problem-dependent multiplier $m$ applied to $ln(Q)$, which adjusts the rate of decay. The parameter $m$ is user-defined, must be a positive real number, and may take values less than 1 for highly ambiguous problems (e.g., open-ended research questions or real-world dynamic systems). In this manner, index $C$ is formulated as:
\begin{equation}
    C = \frac{1}{1+m \cdot ln(Q)}
    \label{eq:c}
\end{equation}
For the present work, parameter $m$ will be set to 1.

To formulate an index for Accuracy, we must determine how closely the answer obtained by the framework aligns with the truth. In the case of numerical predictions in code-generating tasks, this can be quantified through MAPE calculation. The MAPEs of all predicted variables may be averaged to indicate the average relative error. If (MAPE)$_{av}$ represents the averaged MAPE from all predictions, the index $A$ for Accuracy can be expressed as: 
\begin{equation}
    A = 1 - \frac{\text{(MAPE)}_{av}}{100}
    \label{eq:a}
\end{equation}
If MAPE exceeds 100\%, it is set to 100, and index A is 0 (indicating the poorest performance).

Next, as we progress to the third step of the INFINITE methodology, index InI can be defined through a weighted average of the three indices $E$, $C$, and $A$ as:
\begin{equation}
    \text{InI} = w_E \cdot E + w_C \cdot C + w_A \cdot A
    \label{eq:ini}
\end{equation}
where the weights $w_E, w_C$ and $w_A$ are assigned based on the significance of each of the three components, as determined by the user. For the specific problem we are addressing, we will carry out simple averaging and set all weights equal to $\frac{1}{3}$.

The proposed index is a unique tool that comprehensively evaluates three key components of inference: Efficiency, Consistency, and Accuracy. These components are quantified through distinct indices, incorporating factors such as the number of questions asked, the number of queries required to obtain a valid answer (including SB responses), average response time, and comparing results with the ground truth. By integrating these diverse measures, InI provides a holistic assessment of the LLM framework's performance, ensuring a balanced evaluation of its overall capability to answer posed questions.

\section{LSTM model and Forecasting}
\label{sec:lstm}

Standard RNNs \cite{YADAV2024} often struggle with long-term dependencies due to issues such as vanishing gradients. LSTM networks, an advanced RNN variant, address these challenges by incorporating memory cells and gating mechanisms that regulate the flow of information over time \cite{hochreiter1997long}. Details about the LSTM architecture can be found in the literature \cite{gers2000learning}. We have chosen the LSTM model as the coding example to assess OpenAI frameworks. 

The data preprocessing pipeline involves normalizing the input data to a specific range. A sliding window approach is employed to transform raw time series data into fixed-length sequences suitable for input into the LSTM \cite{chen2022forecast}. The model is trained using the backpropagation through time (BPTT) algorithm, optimizing the mean squared error loss function. The trained model is subsequently evaluated on an unseen test set with performance metrics such as Mean Squared Error (MSE), Mean Absolute Error (MAE), Mean Bias (MB), Mean Absolute Percentage Error (MAPE), Mean Fractional Error (MFE), Mean Fractional Bias (MFB), R$^2$ score, and Pearson correlation coefficient $r$ \cite{lane2003introduction, BOYLAN20064946, sedgwick2012pearson}.

The authors developed an LSTM model (LSTM-H) and tested it to identify the optimal settings for forecasting meteorological variables. These settings are presented in Table~\ref{tab:settings}. GPT, OAI1, and OAI3 were instructed to construct an LSTM model based on these settings.

\begin{table}[h]
    \centering
    \caption{LSTM-H model optimal settings for forecasting meteorological variables. (Tr-Te) denotes Training-Testing data split}
  
    \begin{tabular}{cccccccc}
    \toprule
         Units & Acti- & Opti- & Batch & Epochs & Data & Data &  Time\\
         & vation & mizer & Size & & Scaling & Split & Steps\\
         & & & & & & (Tr-Te) & \\
        \midrule
        10 & ReLU & Adam & 16 & 10 & MinMax  & 90-10 & 3\\
        & & & & &Scaler & & \\
        \bottomrule
    \end{tabular}
 
    \label{tab:settings}
\end{table}

It is essential to note that the results presented correspond to the final 10\% of the data set, equating to 4,821 ten-minute intervals (approximately 33.3 days). Predictions are denormalized to the original scale for interpretability. For all LSTM models utilized, predictions are visualized alongside observations to evaluate performance.

\section{Description of the Data Set}\label{sec:data}

\begin{figure}[htbp]
    \centering
    \includegraphics[width=1.0\textwidth]{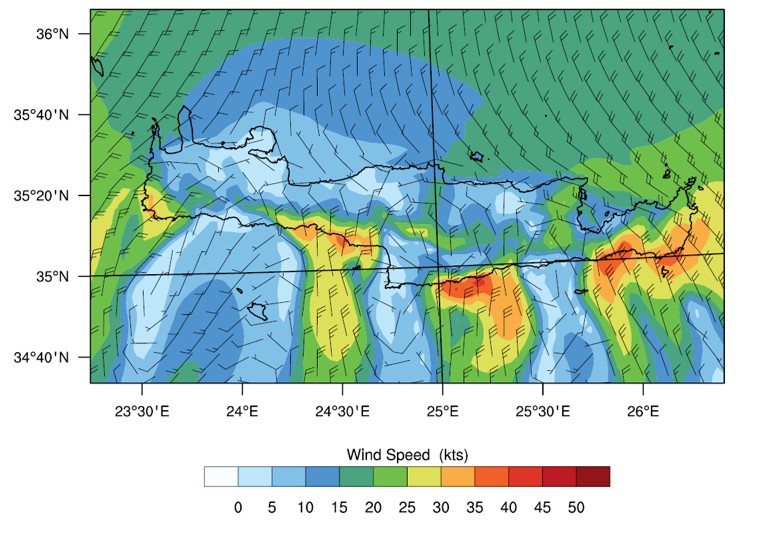}
    \caption{The wind velocity field over Crete, as predicted by WRF for 00:00 GMT on 27 October 2019, is illustrated using both colors (indicating magnitude) and arrows that represent wind direction and strength. The shaft of the arrow points in the direction of the wind, while the barbs denote magnitude (short for 5 knots and long for 10 knots) and are positioned on the side of the shaft from which the wind originates knots.}
    \label{fig:wrf}
\end{figure}

The data set used in this study was collected by a weather station operated by Earth Networks \cite{earthnetworks}, situated at latitude 35.3069$^o$ N and longitude 25.0819$^o$ E, at an elevation of 324 m above sea level. This station serves as a robust source of high-resolution meteorological data. The data set covers precisely one year, from 6 November 2018 at 18:38 local time to 8 November 2019 at 15:19 local time, with measurements taken every 10 minutes, providing a sequential and continuous time series ideal for research purposes. This time window was selected as it was the only period during which continuous time series data could be retrieved without missing values. The station records parameters such as the average temperature over 10 minutes ($^o$C), the lowest temperature during that interval, the highest temperature within the same timeframe (both in $^o$C), dew point temperature ($^o$C), wet bulb temperature ($^o$C), relative humidity (\%), atmospheric pressure at mean sea level (hPa), accumulated daily rainfall resetting at the end of each day (mm), wind speed (m/s), and wind direction (degrees). The wind direction has been analyzed in sine and cosine values to prevent discontinuities at 0$^o$/360$^o$ and to maintain angular relationships for smoother mathematical operations in our machine learning models. A typical wind field for the island of Crete, where the weather station is located, is depicted in Figure~\ref{fig:wrf} for a date included in the dataset. The flow field was generated using WRF numerical weather prediction model simulations 
\cite{7056756}.

The sequential nature of the data makes it particularly suitable for time series analysis, enabling the examination of phenomena such as temperature fluctuations, rainfall events, and wind dynamics. A key strength of this data set lies in its accuracy and reliability, stemming from Earth Networks' established expertise in weather monitoring systems. The high temporal granularity of the data is invaluable for research into local atmospheric processes and both short- and long-term predictions.


\section{Results and Discussion}
\label{sec:preds}
The problem and the procedure adopted are described below. The aim is to generate LSTM models using GPT, OAI1, and OAI3, which simultaneously predict short-term fluctuations in temperature, relative humidity, and wind speed. These variables were selected due to their significance and differing levels of variability, ranging from relatively stable (temperature) to highly dynamic (wind speed). Accurate short-term forecasting of these parameters is essential across multiple domains. For instance, the Canadian Fire Weather Index (FWI), a widely used model for assessing wildfire risk, relies on these parameters 
\cite{wagner1987development, wotton2009interpreting}. Public health also benefits from forecasting these variables, as temperature and humidity affect the heat index, while wind plays a role in air quality by dispersing pollutants and airborne pathogens \cite{10.1063/5.0222192, 10.1063/5.0246654}. 

Initially, the CSV file containing the data was uploaded. The columns labelled "temp", "hum", and "windvel" represent the time series of the variables we wish to predict (i.e., temperature, relative humidity, and wind speed, respectively). We began by asking the same question to all frameworks:

"I have uploaded a file with a continuous time series from a weather station. Retain the first 90\% of the data for training and validation and the last 10\% for testing, in order to train an LSTM model that predicts the variables temp, hum and wind-vel. Write the code in Python that executes this task, including the appropriate error metrics, such as mean squared error, mean absolute error, mean relative error, mean fractional error, mean fractional bias, mean bias, R2 score and Pearson correlation coefficient. Include a graph that compares the last 10\% of the data during the testing phase. The LSTM has one layer with 10 units, a batch size 16, the activation function is ReLU, and the optimiser is Adam with a learning rate of 0.001, running for 10 epochs. It also uses a timestep of 3. If possible, please plot the results of the comparisons."  

The description of the responses for each framework follows:

\subsection{GPT}
\label{sec:gpt}
The GPT framework initially took 53 s to respond. It generated a code, which read from the data set the three required columns and created a training-testing data set only with these three columns, i.e., it used temperature, relative humidity and wind speed only to predict temperature, relative humidity and wind speed. From the question asked, GPT did not understand that all variables in the file had to be taken as input, so a new question (second attempt) was asked, and this was clarified:

"Please do as before, but take as input to the model all given features from the csv file." 

This time, the model generated a Python code within 3 seconds, considering all variables to predict temperature, relative humidity, and wind speed. The code was executed, and predictions for the three variables were obtained. GPT did not manage to generate code that predicted all requested error metrics. It predicted MSE (0.5668), MAE (0.4735), MAPE (1.71$\times 10^{13}$), R$^2$ Score (0.9405) and Pearson $r$ (0.9997). MAPE is erroneous due to division by a minimal number (probably an observed velocity close to 0). As observed, the GPT-generated LSTM model did not manage to evaluate error metrics for each variable separately; instead, it aggregated results across all selected columns and returned an overall mean value. Hence, we cannot comment on the accuracy of the GPT-produced model based on the calculated error metrics. This issue could be remedied by posing an additional question to GPT to calculate error metrics for each variable separately. We will comment on the model's accuracy and perform detailed comparisons with ground truth for each variable later in this section. In summary, the GPT framework provided an answer after 2 attempts and 2 queries. No SB responses were recorded.

\subsection{OAI1} \label{sec:ds}
It took the model 133 seconds to respond and generate Python code. However, an error occurred during execution because the code attempted to overwrite the Pandas DataFrame with all variables and create a Pandas DataFrame of the same name, which included only the three variables: temperature, relative humidity, and wind speed. A new question (second attempt) was posed, indicating the error and requesting that the model correct it. The model responded to the second query after 113 seconds, but the generated code encountered another error during execution. This time, it complained about non-numeric values in the "low temperature" data set. Consequently, a third question (third attempt) was asked, requesting the model to rectify the error. A response was received after 168 seconds, and the model produced a corrected code. However, similar to the case with GPT, the generated code only accessed the three necessary columns from the data set and created a training-testing data set based solely on these three columns. Therefore, a new question (fourth attempt) was posed to clarify that all variables needed to be used as inputs for the generated code. The question was precisely the same as the one asked of GPT:

"Please do as before, but take as input to the model all given features from the csv file."

This time, the framework generated the correct Python code after 103 seconds, utilizing all inputs to predict the three requested variables. The code was executed and predictions for the three variables were obtained. The model predicted all requested error metrics for each variable separately. The results are tabulated in Table~\ref{tab:oai1}.

\FloatBarrier
\begin{figure}[t]
    \centering
    \includegraphics[width=1.\textwidth]{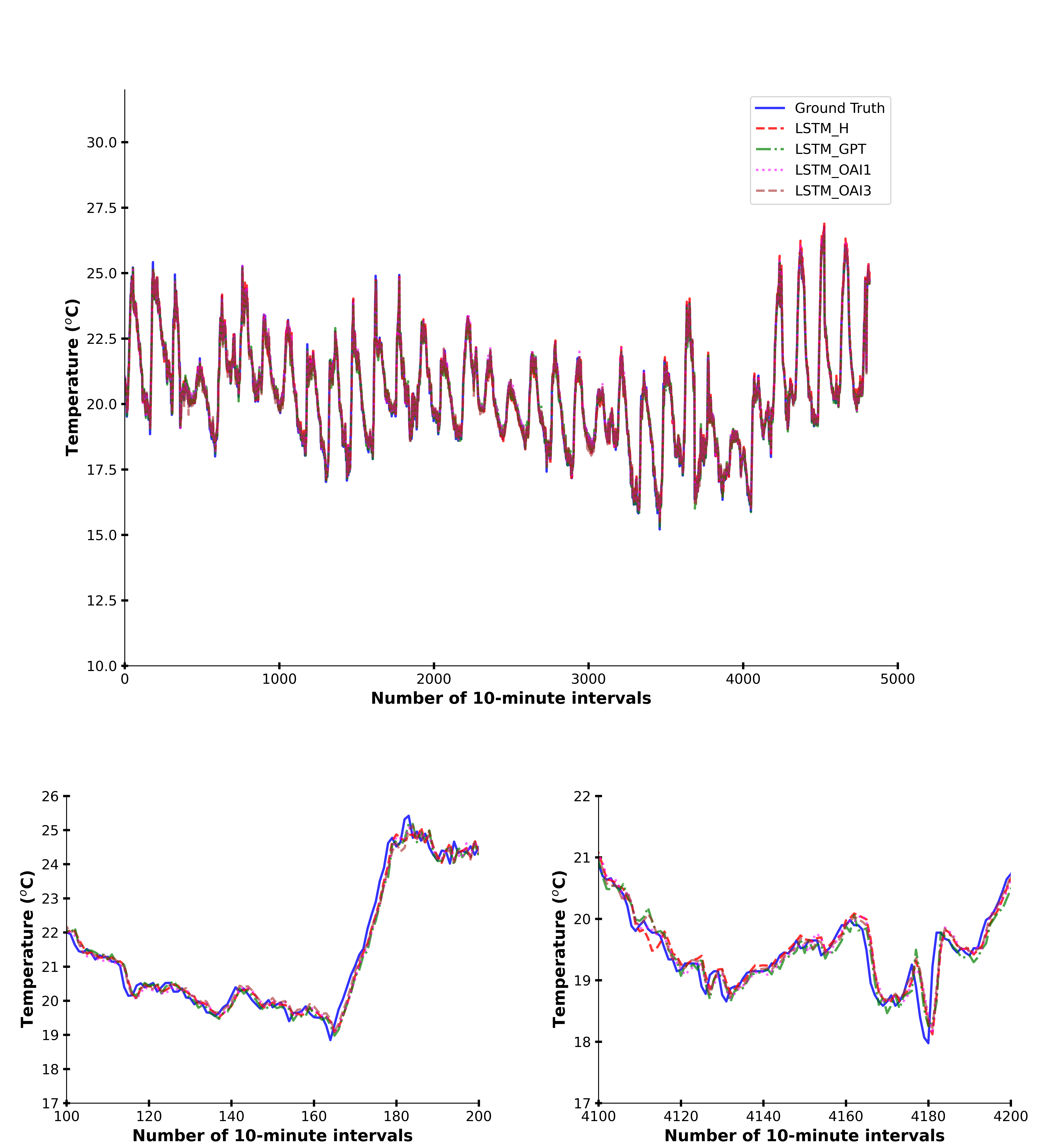}
    \caption{Temperature predictions from all three models and comparisons with ground truth and LSTM-H predictions. LSTM\_H is the model developed by the authors, LSTM\_GPT is the GPT-generated model, LSTM\_OAI1 is the OAI1-generated model, and LSTM\_OAI3 is the OAI3-generated model. The top graph represents the entire testing dataset. The two bottom graphs concentrate on specific time intervals, specifically between 100 and 200 ten-minute intervals (bottom left) and 4100 and 4200 ten-minute intervals (bottom right). This is done to better visualize the difference between the various predictions. }
    \label{fig:temp}
\end{figure}
\FloatBarrier

\begin{table}[h]
    \centering
    \caption{Error metrics for temperature, relative humidity and wind speed, as calculated by the OAI1-generated LSTM model}

    \begin{tabular}{lccc}
    \toprule
         &Temperature& Relative Humidity&Wind Speed \\
       \midrule
      MSE & 0.1214 & 1.1871 & 0.4642\\
      MAE & 0.2838 & 0.7619 & 0.5100\\ 
      MB & -0.2321 & -0.2048 &  -0.1410\\
      MAPE (\%) & 1.3938 & 1.0432 & 9.5998$\times10^8$\\
      MFE (\%) & 1.4161 & 1.0417 & 2.24889$\times10^5$\\ 
      MFB (\%) & -1.1641 & -0.2316 &  -14.1026\\
      R$^2$ & 0.9675 & 0.9861 & 0.8457\\
      Pearson $r$ & 0.9910 &  0.9949 & 0.9235\\
      \bottomrule
    \end{tabular} 
    \label{tab:oai1}
\end{table}

The MAPE and MFE for wind speed are incorrect, likely due to division by a minimal number, close to zero. By examining the error metrics, we can conclude that the OAI1-generated model effectively predicted the three variables satisfactorily. The OAI1 framework delivered a satisfactory response after four attempts and four queries, with no SB responses recorded. Detailed comparisons with the ground truth for each variable will follow.

\subsection{OAI3}
\label{sec:oai3}
 OAI3 responded in 30 seconds and behaved exactly like the GPT framework; it generated code that read the three required columns from the data set and created a training-testing data set consisting solely of these three columns. Specifically, it used temperature, relative humidity, and wind speed values to predict temperature, relative humidity, and wind speed. From the question posed, OAI3 did not grasp that all variables in the file needed to be taken as input, prompting a new question (second attempt) for clarification.

"Please do as before, but take as input to the model all given features from the csv file." 

This time, the framework responded in 21 seconds and generated an LSTM model that considered all variables to predict temperature, relative humidity, and wind speed. The code was executed, and predictions for the three variables were obtained. The model predicted all requested error metrics for each variable individually. The results are presented in Table~\ref{tab:oai3}.
\begin{table}[h]
    \centering
    \caption{Error metrics for temperature, relative humidity and wind speed, as calculated by the OAI3-generated LSTM model}

    \begin{tabular}{lccc}
    \toprule
         &Temperature& Relative Humidity&Wind Speed \\
       \midrule
      MSE & 0.0787 & 1.1518 & 0.6924\\
      MAE & 0.2120 & 0.7428 & 0.6711\\ 
      MB & -0.1433 & -0.1422 &  0.2954\\
      MAPE (\%) & 1.0412 & 1.0238 & 2.0214$\times10^4$\\
      MFE (\%) & 1.0582 & 1.0271 & 34.9757\\ 
      MFB (\%) & -0.7288 & -0.1346 &  29.5465\\
      R$^2$ & 0.9789 & 0.9895 & 0.7698\\
      Pearson $r$ & 0.9923 &  0.9951 & 0.9230\\
      \bottomrule
    \end{tabular} 
    \label{tab:oai3}
\end{table}

The MAPE for wind speed is once again incorrect, likely due to division by a very small number, close to zero. By examining the error metrics for all three variables, we can conclude that the OAI3-generated model also predicted the three variables with reasonable accuracy. The OAI3 framework provided a satisfactory response after two attempts and two queries, with no SB responses recorded. Detailed comparisons with the ground truth for each variable will follow.

\subsection{LSTM-H} \label{sec:lstm-h}
After running the LSTM-H model and obtaining predictions for the test data set for each of the three variables, we compiled Table~\ref{tab:lstm-h} containing all error metrics, specifically MSE, MAE, MB, MAPE, MFE, MFB, R$^2$ score, and the Pearson correlation coefficient $r$. 

\begin{table}[h]
    \centering
    \caption{LSTM-H model error metrics for temperature, relative humidity and wind speed}

    \begin{tabular}{lccc}
    \toprule
         &Temperature& Relative Humidity&Wind Speed \\
       \midrule
      MSE & 0.0994 & 1.4854 & 0.4546\\
      MAE & 0.2366 & 0.8755 & 0.5088\\ 
      MB & 0.1476 & -0.4317 &  0.0890\\
      MAPE (\%) & 1.1569 & 1.1833 & 36.4229\\
      MFE (\%) & 1.1562 & 1.1147 & 19.3069\\ 
      MFB (\%) & 0.7214 & -0.5496 &  3.3800\\
      R$^2$ & 0.9733 & 0.9864 & 0.8488\\
      Pearson $r$ & 0.9901 &  0.9940 & 0.9227\\
      \bottomrule
    \end{tabular} 
    \label{tab:lstm-h}
\end{table}

\begin{figure}[h]
    \centering
    \includegraphics[width=1.\textwidth]{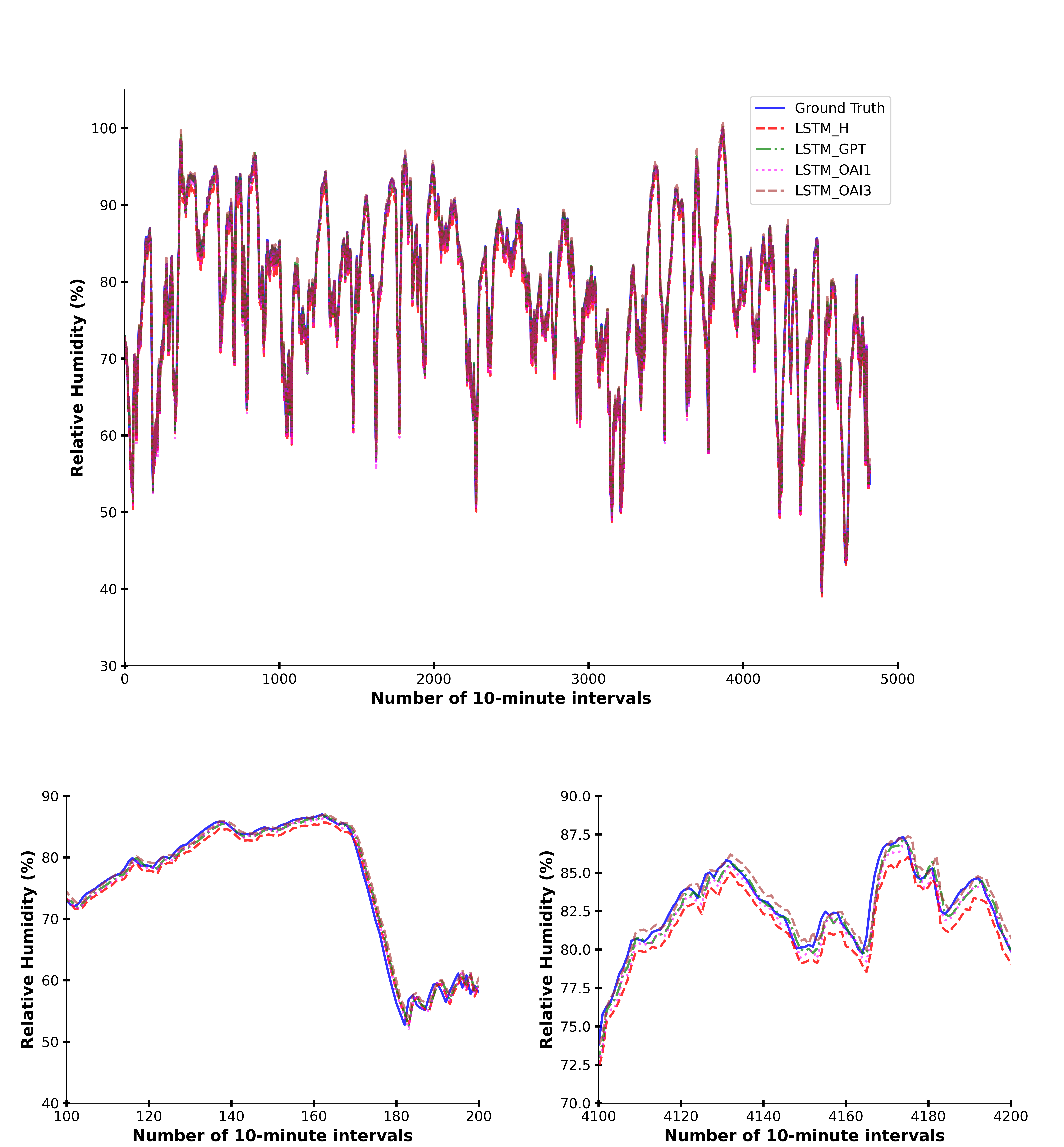}
    \caption{Relative humidity predictions of all three models and comparisons with ground truth and LSTM-H predictions. LSTM\_H is the model developed by the authors, LSTM\_GPT is the GPT-generated model, LSTM\_OAI1 is the OAI1-generated model, and LSTM\_OAI3 is the OAI3-generated model. The top graph is for the whole testing data set. The two bottom graphs focus on specific time intervals, namely between 100 and 200 10-minute intervals (bottom left) and 4100 and 4200 10-minute intervals (bottom right). This is done to better visualize the difference between the different predictions. }
    \label{fig:rh}
\end{figure}

\begin{figure}[h]
    \centering
    \includegraphics[width=1.\textwidth]{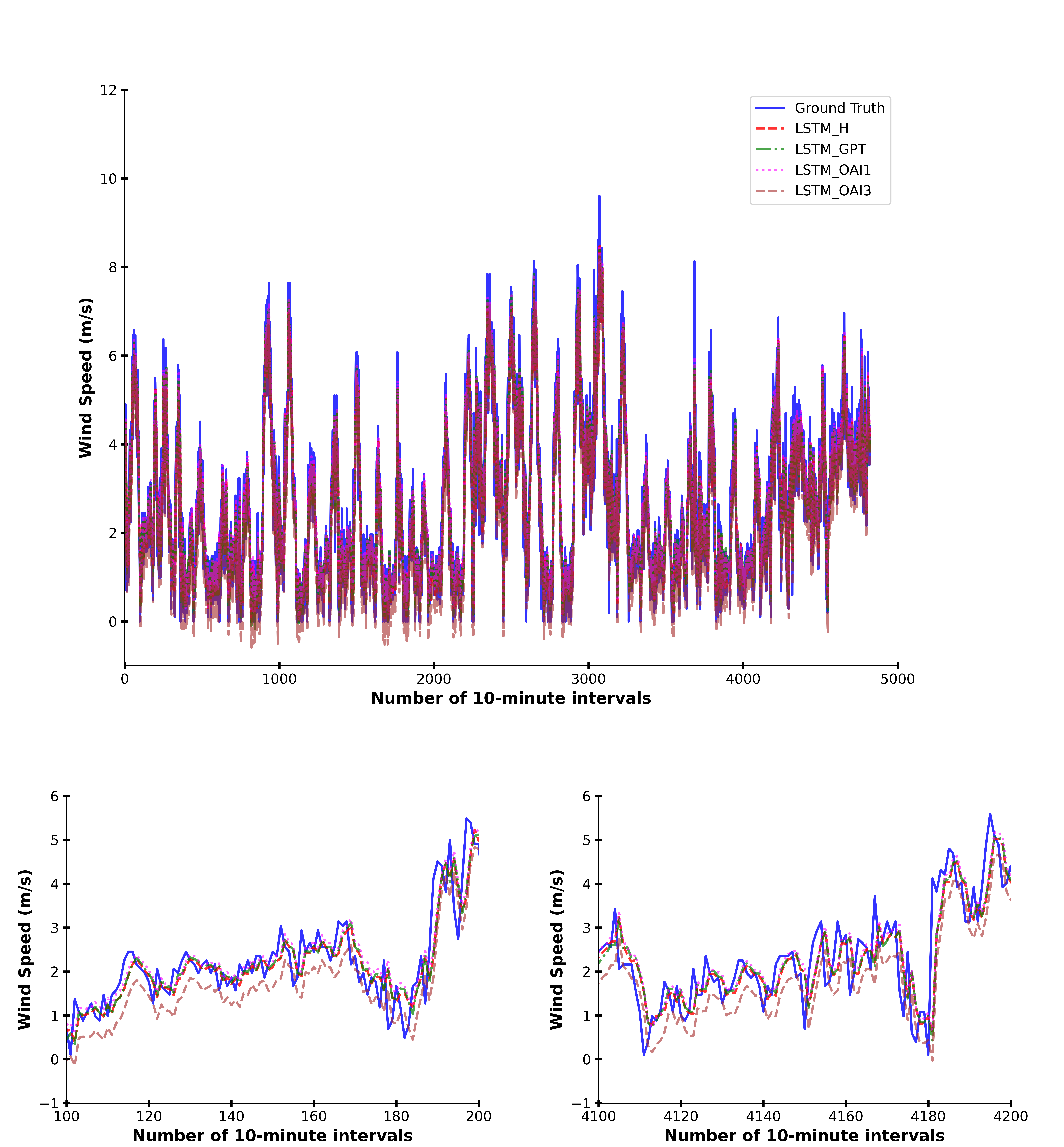}
    \caption{Wind speed predictions of all three models and comparisons with ground truth and LSTM-H predictions. LSTM\_H is the model developed by the authors, LSTM\_GPT is the GPT-generated model, LSTM\_OAI1 is the OAI1-generated model, and LSTM\_OAI3 is the OAI3-generated model.  The top graph is for the whole testing data set. The two bottom graphs focus on specific time intervals, namely between 100 and 200 10-minute intervals (bottom left) and 4100 and 4200 10-minute intervals (bottom right). This is done in order to better visualize the difference between the different predictions. }
    \label{fig:vel}
\end{figure}

The comparisons of the predictions from the LSTM-H model and the three frameworks against observations (i.e., ground truth) are presented in Figures~\ref{fig:temp}, ~\ref{fig:rh}, and ~\ref{fig:vel} for temperature, relative humidity, and wind speed, respectively. All predictions from LSTM models generated by GPT, OAI1, or OAI3 appear to be remarkably close to those of LSTM-H for the testing data set and align well with observations, particularly in the cases of temperature and relative humidity.

\FloatBarrier
\FloatBarrier

As MAPE and R$^2$ scores are reliable metrics for error calculations, we independently calculated the MAPE and R$^2$ scores for the three predicted variables from the GPT-generated model due to their scale independence and interpretability. We also calculated MAPE between the ground truth and wind speed for the OAI1 and OAI3 frameworks, as their calculations were not performed correctly. The results are presented in Tables~\ref{tab:mape} and ~\ref{tab:r2}.

\begin{table}[h]
    \centering
    \caption{MAPE between predictions and ground truth for the LSTM-H, GPT, OAI1, and OAI3 models concerning temperature, relative humidity and wind speed}

    \begin{tabular}{lccc}
    \toprule
        &Temperature& Relative Humidity&Wind Speed \\
       \midrule
      LSTM-H & 1.1569 & 1.1833 & 36.4229\\
      GPT & 0.9026 & 1.0079 & 33.3757\\ 
      OAI1& 1.3938 & 1.0432 & 36.0228\\
      OAI3 & 1.0412 & 1.0238 & 37.4445\\
      \bottomrule
    \end{tabular} 

    \label{tab:mape}
\end{table}

\begin{table}[h]
    \centering
    \caption{R$^2$ score comparisons between predictions and ground truth for the LSTM-H, GPT, OAI1, and OAI3 models concerning temperature, relative humidity and wind speed.}

    \begin{tabular}{lccc}
    \toprule
        &Temperature& Relative Humidity&Wind Speed \\
       \midrule
      LSTM-H & 0.9733 & 0.9864 & 0.8488\\
      GPT & 0.9827 & 0.9892 & 0.85041\\ 
      OAI1& 0.9675 & 0.9861 & 0.8457\\
      OAI3 & 0.9789 & 0.9895 & 0.7698\\
      \bottomrule
    \end{tabular} 

    \label{tab:r2}
\end{table}
As observed in Figures~\ref{fig:temp}, ~\ref{fig:rh} and ~\ref{fig:vel} and the corresponding MAPE and R$^2$ Tables, the predictions of all models are practically indistinguishable and remarkably close to ground truth. When comparing the three OpenAI frameworks (GPT, OAI1 and OAI3), we observe that the GPT framework provides, even if marginally, better accuracy.  It seems that GPT, at present, overtakes in accuracy the other two frameworks for scientific coding tasks. This will be further discussed later in this section.

We will now apply the INFINITE methodology to calculate the InI index and assess the performance of the three LLM frameworks: GPT, OAI1, and OAI3. To identify the parameters that will enable us to measure the performance of each framework, we recorded the number of different questions (attempts) we had to pose to the model until the correct LSTM model was generated, the number of queries submitted each time (whether an SB response or an answer was returned), the response times between a query and a response from the framework and the MAPEs between LSTM predictions and ground truth. For each framework, an average MAPE was calculated from the MAPE values for temperature, relative humidity and wind speed, as shown in Table~\ref{tab:mape}. The main parameters are summarized in Table~\ref{tab:summ}. $ARTpQ$ is calculated using Equation~\ref{eq:art}.

\begin{table}[h]
    \centering
    \caption{Important parameters collected during the execution of the procedure for generating code for an LSTM model using GPT, OAI1 and OAI3.}

    \begin{tabular}{lccccc}
    \toprule
    &Attempts until& Total  &SB & ARTpQ &Average\\
    &correct answer&  queries & responses& (s) & MAPE (\%)\\
    \midrule  
      GPT & 2& 2& 0& 28.53 & 11.76\\ 
      OAI1 & 4 & 4 & 0 &129.25 & 12.81\\
      OAI3 & 2 & 2 & 0 &25.50 & 13.16\\
      \bottomrule
    \end{tabular} 

    \label{tab:summ}
\end{table}

\begin{figure}[h]
    \centering
    \includegraphics[width=1.0\textwidth]{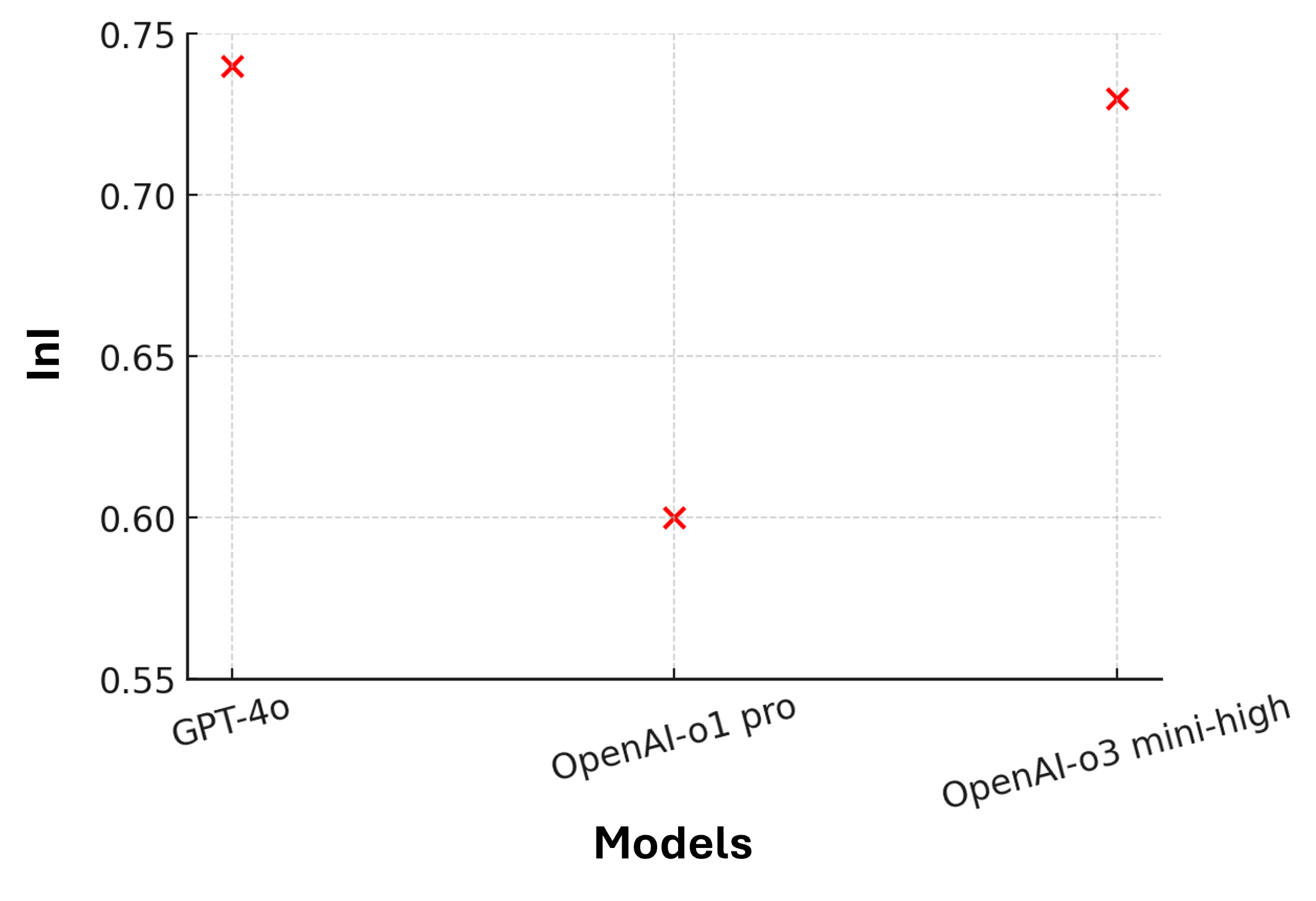}
    \caption{The InI index for the three frameworks assessed for code generation tasks.}
    \label{fig:ini}
\end{figure}

By substituting the values of Table~\ref{tab:summ} into Equations~\ref{eq:esbr}, \ref{eq:eart}, \ref{eq:e}, \ref{eq:c}, \ref{eq:a} and \ref{eq:ini}, we derive Table~\ref{tab:ini}. For better visualization, index InI is shown in Figure~\ref{fig:ini} for the three frameworks.

\begin{table}[h]
    \centering
    \caption{Efficiency, Consistency, Accuracy and InI indices for GPT, OAI1 and OAI3 performance, as determined through the INFINITE methodology.}

    \begin{tabular}{lcccccc}
    \toprule
    &$E_{SBR}$& $E_{ART}$  &$E$ & $C$ &$A$& \textbf{InI}\\
    \midrule
      GPT & 1& 0.5& 0.75& 0.59 & 0.88 & \textbf{0.74}\\ 
      OAI1 & 1 & 0 & 0.50 &0.42 & 0.87 & \textbf{0.60}\\
      OAI3 & 1 & 0.5 & 0.75 &0.59 & 0.86 & \textbf{0.73}\\
      \bottomrule
    \end{tabular} 

    \label{tab:ini}
\end{table}

As observed, GPT surpasses OAI1 at this stage in Efficiency and Consistency, and the two frameworks are very close regarding Accuracy. GPT's performance is also slightly better than OAI3's, in terms of Accuracy. Overall, GPT achieved the highest InI among the three frameworks. It only failed to calculate the requested errors for each variable separately; instead, it aggregated them. Hence, GPT appears to be producing meaningful code for complex machine learning tasks in the shortest time possible. Thus, it seems to be the most appropriate option for scientific coding tasks currently. An extensive literature search yielded no references for this, although there have been limited comparisons between GPT and the newer variants 
\cite{goto2024performance, hu2024can}. We believe this emergent phenomenon occurs because all OpenAI models are regularly fine-tuned based on user interactions, including instances where users correct mistakes or upvote/downvote responses \cite{steiner2024fine}. Since GPT is a public model, it interacts with a more extensive user base and its fine-tuning benefits from a broader range of feedback, potentially enhancing its robustness in certain areas. On the other hand, pay-walled versions, such as OAI1 and OAI3, are fine-tuned based on feedback from their own smaller sets of users. Each model is fine-tuned separately, meaning mistakes corrected in one model do not immediately transfer to another. This indicates that, at least at this stage, GPT performs better than the newer variants of OpenAI in areas where its broader user base has rectified more mistakes. For a comprehensive understanding of the fine-tuning methodologies applicable to LLMs, the interested reader is referred to 
\cite{dodge2020fine}.

\FloatBarrier
\FloatBarrier

\section{Conclusions}\label{sec:conclusions}
In this study, we introduce a new methodology and index to evaluate the performance of LLMs in code-generating tasks. The InI index (calculated through the INFINITE methodology) provides a more comprehensive evaluation by focusing on three key components of inference: Efficiency, Consistency, and Accuracy. Unlike traditional approaches, which often emphasize only accuracy (e.g., MAPE) or response quality, this methodology also incorporates time-based efficiency (average response time and server load) and consistency (the number of iterations needed to reach a correct answer). The InI index results from the combination of indices derived for each key component. This multidimensional approach allows for a more holistic understanding of an LLM’s performance, capturing both the speed and reliability of the model alongside the correctness of the generated code. By incorporating consistency, we account for the model’s learning curve and ability to converge on a solution, which is often overlooked in existing frameworks that primarily focus on outcome precision without addressing model stability. Moreover, by evaluating server load and response times, we provide insights into the scalability and resource efficiency of the framework, making our approach particularly valuable for real-world applications where performance under variable workloads is crucial. This multidimensional assessment offers a richer picture of inference effectiveness and provides actionable insights into both technical and operational aspects of deploying LLMs in code generation tasks. 

Furthermore, we conducted a comparative analysis of OpenAI’s GPT-4o (GPT), OpenAI-o1 pro (OAI1), and OpenAI-o3 mini-high (OAI3) frameworks for generating code in Python to implement an LSTM deep learning model for predicting critical meteorological variables, namely temperature, relative humidity and wind speed. The evaluation focused on performance aspects such as inference efficiency, computational time, forecasting accuracy, memory utilization, and usability. The results indicate that all frameworks can generate functional Python code for LSTM-based forecasting. However, GPT demonstrated a more efficient workflow and outperformed its competitors, particularly OAI1, which required more attempts before producing a fully functional model, indicating less consistency in managing complex programming tasks. OAI3 performed equally well as GPT in terms of efficiency and consistency; however, GPT's predictions showed better accuracy, resulting in a higher overall score. All AI-generated models performed similarly to the manually developed LSTM-H model, with only marginal differences in predictive accuracy. This demonstrates that AI-assisted code generation can yield results comparable to expert-designed implementations, provided that the models are correctly prompted and refined through iterative feedback. 

\FloatBarrier
\FloatBarrier

In summary, GPT's slightly better performance revealed that it remains a competitive framework despite the release of newer variants from OpenAI, namely OAI1 and OAI3. This sustained performance may be attributed to its widespread usage, which allows for extensive fine-tuning based on a larger volume of user interactions, feedback, and error corrections. As a result, GPT benefits from a broader range of refinements, potentially offsetting advancements in newer, more specialized models. The findings of this work highlight the evolving role of AI in computational modeling and scientific research, where the ability to generate accurate and efficient code has significant implications for accelerating discovery and innovation. Future research should explore further refinements in AI-assisted coding, assess additional AI frameworks, and investigate the potential for hybrid approaches that utilize multiple models to optimize performance. Moreover, the InI index may be expanded to incorporate, for instance, metrics such as BLEU to be employed in assessing LLM performance for a broader range of tasks, e.g., text summarization and language translation, among others.


\section*{Declarations}

\begin{itemize}
\item Funding: Not applicable
\item Conflict of interest: All authors declare that they have no conflicts of interest.
\item Ethics approval and consent to participate: Not applicable
\item Consent for publication: Not applicable
\item Data/Code availability: The data set and AI-generated codes, along with the comparison graphs they produced may be downloaded from: https://github.com/nchrkis/infinite 
\item Author contribution: Both authors contributed equally to this work.

\end{itemize}

\noindent



\bigskip

\begin{thebibliography}{10}

\bibitem{openai2022optimizing}
C.~OpenAI.
\newblock Optimizing language models for dialogue.
\newblock \url{https://openai.com/blog/chatgpt}, 2022.
\newblock Accessed in 2023.

\bibitem{openai2023gpt4}
Josh Achiam, Steven Adler, Sandhini Agarwal, Lama Ahmad, Ilge Akkaya, Florencia~Leoni Aleman, Diogo Almeida, Janko Altenschmidt, Sam Altman, Shyamal Anadkat, et~al.
\newblock Gpt-4 technical report.
\newblock {\em arXiv preprint arXiv:2303.08774}, 2023.

\bibitem{translation1}
Ashish Vaswani, Noam Shazeer, Niki Parmar, Jakob Uszkoreit, Llion Jones, Aidan~N Gomez, \L~ukasz Kaiser, and Illia Polosukhin.
\newblock Attention is all you need.
\newblock In I.~Guyon, U.~Von Luxburg, S.~Bengio, H.~Wallach, R.~Fergus, S.~Vishwanathan, and R.~Garnett, editors, {\em Advances in Neural Information Processing Systems}, volume~30, 2017.

\bibitem{translation2}
Mike Lewis, Yinhan Liu, Naman Goyal, Marjan Ghazvininejad, Abdelrahman Mohamed, Omer Levy, Veselin Stoyanov, and Luke Zettlemoyer.
\newblock {BART}: Denoising sequence-to-sequence pre-training for natural language generation, translation, and comprehension.
\newblock In {\em Proceedings of the 58th Annual Meeting of the Association for Computational Linguistics}, pages 7871--7880, Online, July 2020. Association for Computational Linguistics.

\bibitem{speech1}
Dario Amodei, Sundaram Ananthanarayanan, Rishita Anubhai, Jingliang Bai, Eric Battenberg, Carl Case, Jared Casper, Bryan Catanzaro, Qiang Cheng, Guoliang Chen, Jie Chen, Jingdong Chen, Zhijie Chen, Mike Chrzanowski, Adam Coates, Greg Diamos, Ke~Ding, Niandong Du, Erich Elsen, Jesse Engel, Weiwei Fang, Linxi Fan, Christopher Fougner, Liang Gao, Caixia Gong, Awni Hannun, Tony Han, Lappi Johannes, Bing Jiang, Cai Ju, Billy Jun, Patrick LeGresley, Libby Lin, Junjie Liu, Yang Liu, Weigao Li, Xiangang Li, Dongpeng Ma, Sharan Narang, Andrew Ng, Sherjil Ozair, Yiping Peng, Ryan Prenger, Sheng Qian, Zongfeng Quan, Jonathan Raiman, Vinay Rao, Sanjeev Satheesh, David Seetapun, Shubho Sengupta, Kavya Srinet, Anuroop Sriram, Haiyuan Tang, Liliang Tang, Chong Wang, Jidong Wang, Kaifu Wang, Yi~Wang, Zhijian Wang, Zhiqian Wang, Shuang Wu, Likai Wei, Bo~Xiao, Wen Xie, Yan Xie, Dani Yogatama, Bin Yuan, Jun Zhan, and Zhenyao Zhu.
\newblock Deep speech 2 : End-to-end speech recognition in english and mandarin.
\newblock In Maria~Florina Balcan and Kilian~Q. Weinberger, editors, {\em Proceedings of The 33rd International Conference on Machine Learning}, volume~48 of {\em Proceedings of Machine Learning Research}, pages 173--182, New York, New York, USA, 20--22 Jun 2016. PMLR.

\bibitem{speech2}
Anmol Gulati, James Qin, Chung-Cheng Chiu, Niki Parmar, Yu~Zhang, Jiahui Yu, Wei Han, Shibo Wang, Zhengdong Zhang, Yonghui Wu, et~al.
\newblock Conformer: Convolution-augmented transformer for speech recognition.
\newblock {\em arXiv preprint arXiv:2005.08100}, 2020.

\bibitem{bertpaper}
Jacob Devlin, Ming-Wei Chang, Kenton Lee, and Kristina Toutanova.
\newblock {BERT}: Pre-training of deep bidirectional transformers for language understanding.
\newblock In Jill Burstein, Christy Doran, and Thamar Solorio, editors, {\em Proceedings of the 2019 Conference of the North {A}merican Chapter of the Association for Computational Linguistics: Human Language Technologies, Volume 1 (Long and Short Papers)}, pages 4171--4186, Minneapolis, Minnesota, June 2019. Association for Computational Linguistics.

\bibitem{Drikakis:2024_ML3}
Dimitris Drikakis, Ioannis~William Kokkinakis, Daryl Fung, and S.~Michael Spottswood.
\newblock {Self-supervised transformers for turbulent flow time series}.
\newblock {\em Physics of Fluids}, 36(6):065113, 06 2024.

\bibitem{Drikakis:2024_ML2}
Dimitris Drikakis, Ioannis~William Kokkinakis, Daryl Fung, and S.~Michael Spottswood.
\newblock Generalizability of transformer-based deep learning for multidimensional turbulent flow data.
\newblock {\em Physics of Fluids}, 36(2):026102, 02 2024.

\bibitem{huang2023bias}
Dong Huang, Qingwen Bu, Jie Zhang, Xiaofei Xie, Junjie Chen, and Heming Cui.
\newblock Bias assessment and mitigation in llm-based code generation.
\newblock {\em arXiv preprint arXiv:2309.14345}, 2023.

\bibitem{NEURIPS2023_ae9500c4}
Yue Yu, Yuchen Zhuang, Jieyu Zhang, Yu~Meng, Alexander~J Ratner, Ranjay Krishna, Jiaming Shen, and Chao Zhang.
\newblock Large language model as attributed training data generator: A tale of diversity and bias.
\newblock In A.~Oh, T.~Naumann, A.~Globerson, K.~Saenko, M.~Hardt, and S.~Levine, editors, {\em Advances in Neural Information Processing Systems}, volume~36, pages 55734--55784, NY 12571, USA, 2023. {Curran Associates, Inc.}
\newblock \url{https:////proceedings.neurips.cc/paper_files/paper/2023/file/ae9500c4f5607caf2eff033c67daa9d7-Paper-Datasets_and_Benchmarks.pdf}.

\bibitem{10.1145/3637528.3671458}
Sunhao Dai, Chen Xu, Shicheng Xu, Liang Pang, Zhenhua Dong, and Jun Xu.
\newblock Bias and unfairness in information retrieval systems: New challenges in the llm era.
\newblock In {\em Proceedings of the 30th ACM SIGKDD Conference on Knowledge Discovery and Data Mining}, KDD '24, page 6437–6447, New York, NY, USA, 2024. Association for Computing Machinery.

\bibitem{jin2023time}
Ming Jin, Shiyu Wang, Lintao Ma, Zhixuan Chu, James~Y Zhang, Xiaoming Shi, Pin-Yu Chen, Yuxuan Liang, Yuan-Fang Li, Shirui Pan, et~al.
\newblock Time-llm: Time series forecasting by reprogramming large language models.
\newblock {\em arXiv preprint arXiv:2310.01728}, 2023.

\bibitem{yu2023temporal}
Xinli Yu, Zheng Chen, Yuan Ling, Shujing Dong, Zongyi Liu, and Yanbin Lu.
\newblock Temporal data meets llm--explainable financial time series forecasting.
\newblock {\em arXiv preprint arXiv:2306.11025}, 2023.

\bibitem{zhang2024scaling}
Biao Zhang, Zhongtao Liu, Colin Cherry, and Orhan Firat.
\newblock When scaling meets llm finetuning: The effect of data, model and finetuning method.
\newblock {\em arXiv preprint arXiv:2402.17193}, 2024.

\bibitem{ajwani2024llm}
Rohan Ajwani, Shashidhar~Reddy Javaji, Frank Rudzicz, and Zining Zhu.
\newblock Llm-generated black-box explanations can be adversarially helpful.
\newblock {\em arXiv preprint arXiv:2405.06800}, 2024.

\bibitem{nam2024using}
Daye Nam, Andrew Macvean, Vincent Hellendoorn, Bogdan Vasilescu, and Brad Myers.
\newblock Using an llm to help with code understanding.
\newblock In {\em Proceedings of the IEEE/ACM 46th International Conference on Software Engineering}, pages 1--13, 2024.

\bibitem{liu2024exploring}
Fang Liu, Yang Liu, Lin Shi, Houkun Huang, Ruifeng Wang, Zhen Yang, Li~Zhang, Zhongqi Li, and Yuchi Ma.
\newblock Exploring and evaluating hallucinations in llm-powered code generation.
\newblock {\em arXiv preprint arXiv:2404.00971}, 2024.

\bibitem{tian2023chatgpt}
Haoye Tian, Weiqi Lu, Tsz~On Li, Xunzhu Tang, Shing-Chi Cheung, Jacques Klein, and Tegawend{\'e}~F Bissyand{\'e}.
\newblock Is chatgpt the ultimate programming assistant--how far is it?
\newblock {\em arXiv preprint arXiv:2304.11938}, 2023.

\bibitem{liu2024refining}
Yue Liu, Thanh Le-Cong, Ratnadira Widyasari, Chakkrit Tantithamthavorn, Li~Li, Xuan-Bach~D Le, and David Lo.
\newblock Refining chatgpt-generated code: Characterizing and mitigating code quality issues.
\newblock {\em ACM Transactions on Software Engineering and Methodology}, 33(5):1--26, 2024.

\bibitem{NEURIPS2022_8bb0d291}
Takeshi Kojima, Shixiang~(Shane) Gu, Machel Reid, Yutaka Matsuo, and Yusuke Iwasawa.
\newblock Large language models are zero-shot reasoners.
\newblock In S.~Koyejo, S.~Mohamed, A.~Agarwal, D.~Belgrave, K.~Cho, and A.~Oh, editors, {\em Advances in Neural Information Processing Systems}, volume~35, pages 22199--22213, NY 12571, USA, 2022. Curran Associates, Inc.
\newblock \url{https://proceedings.neurips.cc/paper_files/paper/2022/file/8bb0d291acd4acf06ef112099c16f326-Paper-Conference.pdf}.

\bibitem{zhang2022automatic}
Zhuosheng Zhang, Aston Zhang, Mu~Li, and Alex Smola.
\newblock Automatic chain of thought prompting in large language models.
\newblock {\em arXiv preprint arXiv:2210.03493}, 2022.

\bibitem{jin2024impact}
Mingyu Jin, Qinkai Yu, Dong Shu, Haiyan Zhao, Wenyue Hua, Yanda Meng, Yongfeng Zhang, and Mengnan Du.
\newblock The impact of reasoning step length on large language models.
\newblock {\em arXiv preprint arXiv:2401.04925}, 2024.

\bibitem{8675201}
Carole-Jean Wu, David Brooks, Kevin Chen, Douglas Chen, Sy~Choudhury, Marat Dukhan, Kim Hazelwood, Eldad Isaac, Yangqing Jia, Bill Jia, Tommer Leyvand, Hao Lu, Yang Lu, Lin Qiao, Brandon Reagen, Joe Spisak, Fei Sun, Andrew Tulloch, Peter Vajda, Xiaodong Wang, Yanghan Wang, Bram Wasti, Yiming Wu, Ran Xian, Sungjoo Yoo, and Peizhao Zhang.
\newblock Machine learning at facebook: Understanding inference at the edge.
\newblock In {\em 2019 IEEE International Symposium on High Performance Computer Architecture (HPCA)}, pages 331--344, 2019.

\bibitem{10.1145/3183713.3183751}
Yao Lu, Aakanksha Chowdhery, Srikanth Kandula, and Surajit Chaudhuri.
\newblock Accelerating machine learning inference with probabilistic predicates.
\newblock In {\em Proceedings of the 2018 International Conference on Management of Data}, SIGMOD '18, page 1493–1508, New York, NY, USA, 2018. Association for Computing Machinery.

\bibitem{creswell2022selection}
Antonia Creswell, Murray Shanahan, and Irina Higgins.
\newblock Selection-inference: Exploiting large language models for interpretable logical reasoning.
\newblock {\em arXiv preprint arXiv:2205.09712}, 2022.

\bibitem{pmlr-v202-sheng23a}
Ying Sheng, Lianmin Zheng, Binhang Yuan, Zhuohan Li, Max Ryabinin, Beidi Chen, Percy Liang, Christopher Re, Ion Stoica, and Ce~Zhang.
\newblock {F}lex{G}en: High-throughput generative inference of large language models with a single {GPU}.
\newblock In Andreas Krause, Emma Brunskill, Kyunghyun Cho, Barbara Engelhardt, Sivan Sabato, and Jonathan Scarlett, editors, {\em Proceedings of the 40th International Conference on Machine Learning}, volume 202 of {\em Proceedings of Machine Learning Research}, pages 31094--31116, 23--29 Jul 2023.

\bibitem{chitty2024llm}
Krishna~Teja Chitty-Venkata, Siddhisanket Raskar, Bharat Kale, Farah Ferdaus, Aditya Tanikanti, Ken Raffenetti, Valerie Taylor, Murali Emani, and Venkatram Vishwanath.
\newblock Llm-inference-bench: Inference benchmarking of large language models on ai accelerators.
\newblock In {\em SC24-W: Workshops of the International Conference for High Performance Computing, Networking, Storage and Analysis}, pages 1362--1379. IEEE, 2024.

\bibitem{lukasik2024metric}
Michal Lukasik, Harikrishna Narasimhan, Aditya~Krishna Menon, Felix Yu, and Sanjiv Kumar.
\newblock Metric-aware llm inference.
\newblock {\em arXiv preprint arXiv:2403.04182}, 2024.

\bibitem{he2024uellm}
Yiyuan He, Minxian Xu, Jingfeng Wu, Wanyi Zheng, Kejiang Ye, and Chengzhong Xu.
\newblock Uellm: A unified and efficient approach for llm inference serving.
\newblock {\em arXiv preprint arXiv:2409.14961}, 2024.

\bibitem{MARSILI20221}
Matteo Marsili and Yasser Roudi.
\newblock Quantifying relevance in learning and inference.
\newblock {\em Physics Reports}, 963:1--43, 2022.

\bibitem{PSAROS2023111902}
Apostolos~F. Psaros, Xuhui Meng, Zongren Zou, Ling Guo, and George~Em Karniadakis.
\newblock Uncertainty quantification in scientific machine learning: Methods, metrics, and comparisons.
\newblock {\em Journal of Computational Physics}, 477:111902, 2023.

\bibitem{hochreiter1997long}
Sepp Hochreiter and J\"{u}rgen Schmidhuber.
\newblock Long short-term memory.
\newblock {\em Neural Comput.}, 9(8):1735–1780, November 1997.

\bibitem{liu2024largelanguagemodelscausal}
Xiaoyu Liu, Paiheng Xu, Junda Wu, Jiaxin Yuan, Yifan Yang, Yuhang Zhou, Fuxiao Liu, Tianrui Guan, Haoliang Wang, Tong Yu, Julian McAuley, Wei Ai, and Furong Huang.
\newblock Large language models and causal inference in collaboration: A comprehensive survey, 2024.

\bibitem{pearl1998graphs}
Judea Pearl.
\newblock Graphs, causality, and structural equation models.
\newblock {\em Sociological Methods \& Research}, 27(2):226--284, 1998.

\bibitem{pearl2009causality}
Judea Pearl.
\newblock Graphical models for probabilistic and causal reasoning.
\newblock In Philippe Smets, editor, {\em Quantified Representation of Uncertainty and Imprecision}, pages 367--389. Springer Netherlands, Dordrecht, 1998.

\bibitem{zeng2022surveycausalinferenceframeworks}
Jingying Zeng and Run Wang.
\newblock A survey of causal inference frameworks.
\newblock {\em arXiv preprint arXiv:2209.00869}, 2022.

\bibitem{ji2023benchmarking}
Zhenlan Ji, Pingchuan Ma, Zongjie Li, and Shuai Wang.
\newblock Benchmarking and explaining large language model-based code generation: A causality-centric approach.
\newblock {\em arXiv preprint arXiv:2310.06680}, 2023.

\bibitem{vu2021scnet}
Thang Vu, Haeyong Kang, and Chang~D Yoo.
\newblock Scnet: Training inference sample consistency for instance segmentation.
\newblock In {\em Proceedings of the AAAI Conference on Artificial Intelligence}, volume~35, pages 2701--2709, 2021.

\bibitem{mitchell2022enhancing}
Eric Mitchell, Joseph~J Noh, Siyan Li, William~S Armstrong, Ananth Agarwal, Patrick Liu, Chelsea Finn, and Christopher~D Manning.
\newblock Enhancing self-consistency and performance of pre-trained language models through natural language inference.
\newblock {\em arXiv preprint arXiv:2211.11875}, 2022.

\bibitem{10769165}
Ioanna Valsamara, Christos Papaioannidis, and Ioannis Pitas.
\newblock Efficient data utilization in deep neural networks for inference reliability.
\newblock In {\em 2024 IEEE International Conference on Image Processing Challenges and Workshops (ICIPCW)}, pages 4142--4147, 2024.

\bibitem{papineni2002bleu}
Kishore Papineni, Salim Roukos, Todd Ward, and Wei-Jing Zhu.
\newblock Bleu: a method for automatic evaluation of machine translation.
\newblock In {\em Proceedings of the 40th annual meeting of the Association for Computational Linguistics}, pages 311--318, 2002.

\bibitem{ren2020codebleu}
Shuo Ren, Daya Guo, Shuai Lu, Long Zhou, Shujie Liu, Duyu Tang, Neel Sundaresan, Ming Zhou, Ambrosio Blanco, and Shuai Ma.
\newblock Codebleu: a method for automatic evaluation of code synthesis.
\newblock {\em arXiv preprint arXiv:2009.10297}, 2020.

\bibitem{YADAV2024}
Hemant Yadav and Amit Thakkar.
\newblock Noa-lstm: An efficient lstm cell architecture for time series forecasting.
\newblock {\em Expert Systems with Applications}, 238:122333, 2024.

\bibitem{gers2000learning}
Felix~A Gers, J{\"u}rgen Schmidhuber, and Fred Cummins.
\newblock Learning to forget: Continual prediction with lstm.
\newblock {\em Neural computation}, 12(10):2451--2471, 2000.

\bibitem{chen2022forecast}
Chengcheng Chen, Qian Zhang, Mahsa~H Kashani, Changhyun Jun, Sayed~M Bateni, Shahab~S Band, Sonam~Sandeep Dash, and Kwok-Wing Chau.
\newblock Forecast of rainfall distribution based on fixed sliding window long short-term memory.
\newblock {\em Engineering Applications of Computational Fluid Mechanics}, 16(1):248--261, 2022.

\bibitem{lane2003introduction}
David~M. Lane.
\newblock Online statistics education: A multimedia course of study.
\newblock \url{http://onlinestatbook.com/}, Accessed 2025.
\newblock Project Leader: David M. Lane, Rice University.

\bibitem{BOYLAN20064946}
James~W. Boylan and Armistead~G. Russell.
\newblock Pm and light extinction model performance metrics, goals, and criteria for three-dimensional air quality models.
\newblock {\em Atmospheric Environment}, 40(26):4946--4959, 2006.
\newblock Special issue on Model Evaluation: Evaluation of Urban and Regional Eulerian Air Quality Models.

\bibitem{sedgwick2012pearson}
Phillip Sedgewick.
\newblock Pearson’s correlation coefficient.
\newblock {\em BMJ}, 345:e4483, 2012.
\newblock Published 04 July 2012.

\bibitem{earthnetworks}
{Earth Networks}.
\newblock Weather data, forecasting, and environmental monitoring solutions.
\newblock \url{https://www.earthnetworks.com}.
\newblock Accessed: 2025-01-17.

\bibitem{7056756}
Nicholas Christakis, Theodoros Katsaounis, George Kossioris, and Michael Plexousakis.
\newblock On the performance of the wrf numerical model over complex terrain on a high performance computing cluster.
\newblock In {\em 2014 IEEE Intl Conf on High Performance Computing and Communications, 2014 IEEE 6th Intl Symp on Cyberspace Safety and Security, 2014 IEEE 11th Intl Conf on Embedded Software and Syst (HPCC,CSS,ICESS)}, pages 298--303, 2014.

\bibitem{wagner1987development}
C.~E. van Wagner.
\newblock Development and structure of the canadian forest fire weather index system.
\newblock Forestry Technical Report~35, Canadian Forestry Service, 1987.
\newblock Ref. 63, CABI Record Number: 19910646918.

\bibitem{wotton2009interpreting}
B~Mike Wotton.
\newblock Interpreting and using outputs from the canadian forest fire danger rating system in research applications.
\newblock {\em Environmental and ecological statistics}, 16:107--131, 2009.

\bibitem{10.1063/5.0222192}
C.~Le~Ribault, I.~Vinkovic, and S.~Simoëns.
\newblock Large eddy simulation of droplet dispersion and deposition over street canyons.
\newblock {\em Physics of Fluids}, 36(11):113313, 11 2024.

\bibitem{10.1063/5.0246654}
George Zodo, Harshavardhan Konka, Svetlana Stevanovic, and Jorg (Jörg~Schlüter) Schluter.
\newblock Simulation of the transition of respiratory droplets to aerosol states: Implications for pathogen spread.
\newblock {\em Physics of Fluids}, 37(1):015188, 01 2025.

\bibitem{goto2024performance}
Hiroki Goto, Yoshioki Shiraishi, and Seiji Okada.
\newblock Performance evaluation of gpt-4o and o1-preview using the certification examination for the japanese'operations chief of radiography with x-rays'.
\newblock {\em Cureus}, 16(11):e74262, 2024.

\bibitem{hu2024can}
Haichuan Hu, Ye~Shang, Guolin Xu, Congqing He, and Quanjun Zhang.
\newblock Can gpt-o1 kill all bugs? an evaluation of gpt-family llms on quixbugs.
\newblock {\em arXiv preprint arXiv:2409.10033}, 2024.

\bibitem{steiner2024fine}
Aaron Steiner, Ralph Peeters, and Christian Bizer.
\newblock Fine-tuning large language models for entity matching.
\newblock {\em arXiv preprint arXiv:2409.08185}, 2024.

\bibitem{dodge2020fine}
Jesse Dodge, Gabriel Ilharco, Roy Schwartz, Ali Farhadi, Hannaneh Hajishirzi, and Noah Smith.
\newblock Fine-tuning pretrained language models: Weight initializations, data orders, and early stopping.
\newblock {\em arXiv preprint arXiv:2002.06305}, 2020.

\end{thebibliography}

\end{document}